\documentclass[aps,prd,twocolumn,groupedaddress]{revtex4-2}

\usepackage{graphicx}
\usepackage{amsmath}
\usepackage{aasms}

\usepackage{booktabs,tabularx}
\usepackage{float}

\usepackage{booktabs}
\usepackage{float}
\usepackage{longtable}

\begin{document}


\title{Identification of GRB precursors in Fermi-GBM bursts}



\author{Paul Coppin}
\email{paulcppn@gmail.com}

\author{Krijn D. de Vries}
\author{Nick van Eijndhoven}
\affiliation{Department of Physics and Astronomy\\ Vrije Universiteit Brussel, Pleinlaan 2, Elsene, Belgium}


\date{\today}

\begin{abstract}
We present an analysis of more than 11 years of Fermi-GBM data in which 217 Gamma-Ray Bursts (GRBs) are found for which their main burst is preceded by a precursor flash. We find  that short GRBs ($<2$~s) are $\sim10$ times less likely to produce a precursor than long GRBs. The quiescent time profile, given by the time between the precursor and the main burst, is well described by a double Gaussian distribution, indicating that the observed precursors have two distinct physical progenitors. The light curves of the identified precursor GRBs are publicly available in an online catalog (\url{https://icecube.wisc.edu/~grbweb_public/Precursors.html}).
\end{abstract}


\maketitle

\section{Introduction}
Gamma-Ray Bursts (GRBs) are cataclysmic transient cosmic events characterized by the emission of one or multiple flashes of gamma radiation. They are the most powerful outbursts of electromagnetic radiation in our universe and a possible source of (ultra) high-energy cosmic rays \cite{245, 246, 247, Zhang}. The duration of GRBs can be described using a bi-modal distribution \cite{244}, indicating the existence of two progenitor source classes. In general, bursts lasting longer than 2~s are related to the collapse of a super-massive star, as confirmed by the observation of type-Ic supernovae in coincidence with long GRBs \cite{212,242,243}. Short bursts, lasting less than 2~s, are believed to occur when two co-orbiting neutron stars collide. Evidence for this model was recently obtained by the detection of gravitational waves from a binary neutron star merger followed by a short GRB \cite{012,013}.

The main outburst of gamma radiation, called the prompt phase, is followed by an afterglow stage in which the ejected matter collides with the surrounding medium. Thanks to multi-wavelength observations, ranging from X-ray to radio, the physical processes related to this afterglow emission are well understood \cite{Zhang}. Apart from the prompt and afterglow phases, there is a third emission phase, called the GRB precursor. Precursors are typically defined as relatively dim gamma-ray flashes that occur before the prompt emission. Previous studies \cite{197,049,107,050,110,109,162,190,ICRCposter,108,203,204,217,218} found that precursor flashes occur in a subset of both long and short GRBs. The fraction of bursts in which a precursor is observed strongly depends on the method and criteria used to define a precursor and typically ranges from 3\% to 20\%.

Numerous models have been proposed to explain precursor flashes and typically apply to a specific class of GRB progenitors. In the case of short GRBs caused by the merger of a binary neutron star system, the interaction between the magnetic fields of the neutron stars could induce a Poynting flux prior to the GRB \cite{142,210}. Alternatively, tidal forces could induce resonance modes that lead to a failure of the neutron star crust, potentially releasing enough energy to be observed as a precursor \cite{250,219,248}. In the case of long GRBs, precursors could be related to the emission of an early weak jet. Such jets are predicted in two-step engines, where the star first collapses to a proto-neutron star or spinar, before collapsing to a black hole \cite{253,254}. Multiple jets could also result from an effective turn-off of the central engine, related to sudden changes in the accretion rate \cite{124,190}. In this case, the precursor and prompt emission would be caused by the same physical processes and thus have similar observational properties. Finally, precursors could also be related to the transition of the GRB ejecta into the optically thin phase \cite{138,154,211,251,252}. The precursor is then typically expected to have a thermal spectrum, as was for instance observed for the first identified precursor \cite{197}. Currently, there is no consensus on the origin of precursor flashes and, most likely, more than one model will be needed to explain all observed precursors. Given that precursors only occur in a subset of all GRBs, an extensive study is thus required to uncover their physical origins.

We performed an automated search that identifies precursor flashes observed by the Fermi-GBM detector. Out of a total sample of 2364 GRBs, 244 precursors were identified originating from 217 GRBs, of which 139 are newly identified GRBs with precursor emission.

In this paper, we present the details of our selection and show that short GRBs are $\sim$10 times less likely to produce a precursor than long GRBs. We performed an analysis on the quiescent time profile, given by the time between the precursor flash and the main burst. The increased statistics from our search allowed us to identify a novel feature in the quiescent time profile, which is well described by a double Gaussian distribution, indicating that the observed precursors have two distinct physical progenitors. To allow for follow-up studies, searching for coincidences with other astrophysical messengers, such as neutrinos and gravitational waves, the obtained results for each individual GRB are presented in the Appendix \ref{sub:cat} 
and have been made available via an online tool~\cite{GRBweb}.

\section{Data}
The Fermi Gamma-ray Space Telescope is currently the most efficient GRB detection satellite in orbit. Its two main instruments are the Large Area Telescope (LAT) and the Gamma-ray Burst Monitor (GBM). Whereas LAT has a sky coverage of 20\%, GBM continuously observes the full region of the sky not occulted by Earth. On average, the GBM and LAT detect 240 and 18 GRBs per year, respectively~\cite{206,158,GBMGRBCatalog}. In this study we analyzed 2684 GRBs, using all GBM recorded bursts up to the year 2020.

The GBM telescope is composed of 12 sodium iodide (NaI) and two bismuth germanate (BGO) detectors. Trigger and localization information is provided by the NaI detectors, which are sensitive to gamma rays of 8~keV to 1~MeV. The BGO detectors, which will not be used in this analysis, are sensitive from 200~keV to 40~MeV and serve to cover the energy gap with the LAT \cite{158}.

The GBM burst data was obtained from the Fermi Science Support Center \cite{GBMPublicData} and provides the raw photon counts as a function of time and energy for each of the 14 detectors. Time-Tagged Event (TTE) data provides the highest temporal resolution of $2\ \mathrm{\mu s}$. Since August 2010, TTE data is available over a time window $\left[t_{tr}-135\ \mathrm{s},\ t_{tr}+300\ \mathrm{s}\right]$, where $t_{tr}$ is the GBM detector trigger time. Before August 2010, TTE data is only available starting 30~s before $t_{tr}$, but again up to 300~s after $t_{tr}$. CTIME data is provided over a 2000~s time window centered around $t_{tr}$, but has a coarser nominal resolution of 0.256~s. To allow the detection of very short emission periods, we have used TTE data whenever available. CTIME data was used to extend the examined time window to 1000~s before and after the trigger time.

\section{Method}
For every burst, we select the GBM NaI detectors that were triggered by the GRB. If more than three detectors were triggered, only the three triggered detectors which were pointing closest to the burst location are used. The data analysis is two-fold. An initial analysis on raw time data is performed to characterize the background, allowing to capture global fluctuations. Subsequently, a Bayesian Block (BB) algorithm~\cite{106} is used to select the physical signal regions.

Our analysis aims at identifying all emission periods in which gamma-ray activity is observed from the detected GRBs. This is achieved by constructing background subtracted light curves. For more than 90\% of the identified bursts, a stable background fit is first obtained between 1000~s and 800~s before $t_{tr}$, marking the start time of the analysis interval. The end time of the analysis interval is set 50~s past the end of the Fermi T90 interval, defined as the central time window that contains 90\% of the fluence of the GRB. If the Fermi T90 exceeds 250~s, we set the end time $0.2\cdot \mathrm{T90}$ instead of 50~s beyond the T90 interval. One final consideration is that a minority of all bursts have one or more gaps in their light curves due to missing data. For those bursts, we only examine the continuous data taking period containing $t_{tr}$. This choice is motivated by the observation that for $<$1\% of all bursts, additional data is available at earlier times.

We automated the selection of background time intervals in which no increased gamma-ray activity is observed. Our selection is therefore fully reproducible and based on physically motivated parameters. Background times are selected based on the requirement that the rate does not undergo a sudden increase. For this purpose, we use an algorithm similar to the Fermi-GBM online trigger \cite{158}, which compares the observed rate to a prediction based on a fit to the rate at earlier times. The rate in the identified background intervals is then extrapolated to intermediate regions. As such, we obtain a fully data driven estimate for the background rate over the full light curve. A more detailed description of this method and a motivation for the use of Poisson statistics are provided in Appendix \ref{sub:bg} and \ref{sub:Poisson}, respectively.

\begin{figure*}[t]
\centering
{\includegraphics[width=\linewidth]{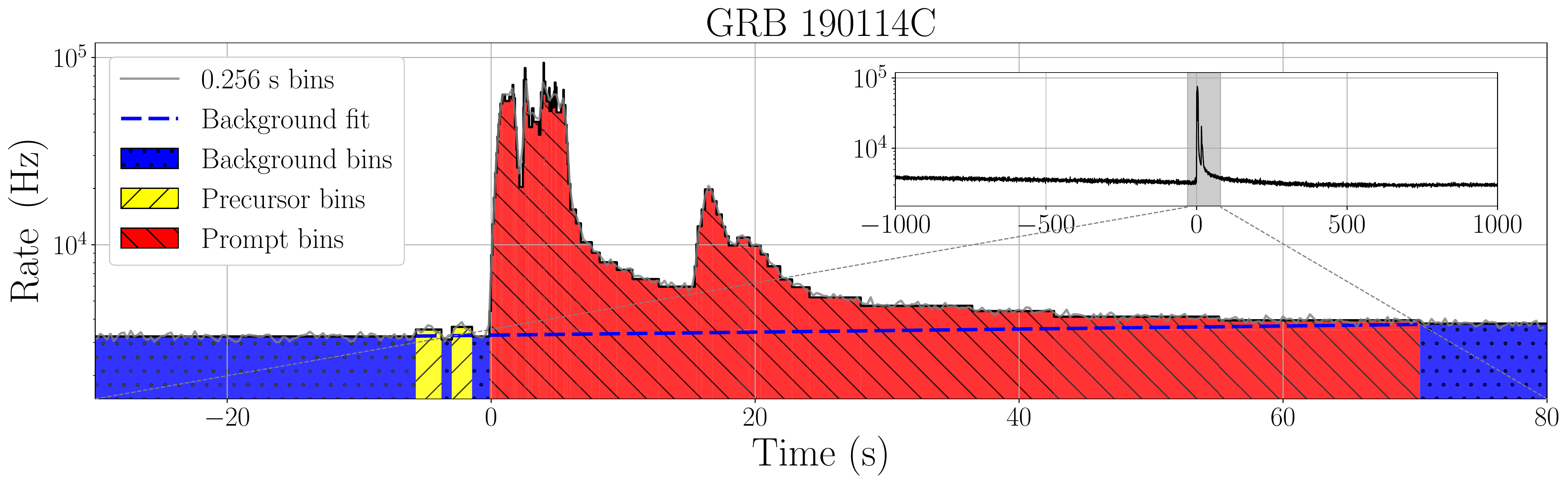}}
\caption{Illustration of the Bayesian block light curve of GRB 190114C. Two dim precursors (yellow) are observed, starting 5.57~s and 2.85~s before the onset of the prompt emission (red). The total photon count in the precursor bins is 6839 and 5590, respectively, exceeding the expected background count by more than 6$\sigma$ in both cases. The time range displayed in this figure corresponds to the grey shaded area in the inset image, which displays the full light curve.}
\label{fig:GRB190114C}
\end{figure*}

Having characterized the background rate, we proceed by rebinning the data using the Bayesian Block (BB) algorithm~\cite{106}. The BB algorithm was specifically designed to identify localized structures, such as bursts, in GRB light curves. It optimizes both the number of bins and the location of the bin edges by maximizing a fitness function. For every selected GBM detector, we construct a BB light curve. In addition, a single BB light curve based on a combination of the photon counts of the selected detectors is also constructed for every burst. These combined light curves contain the largest statistics and will thus serve as the basis for our selection. To illustrate the BB procedure, the light curve of GRB 190114C is displayed in Fig.~\ref{fig:GRB190114C}.

\section{Analysis}
To quantify the physical signal, a background subtraction procedure is applied. The background rate is integrated over each bin to estimate the total number of background events $N_b$. Subtracting $N_b$ from the observed count and dividing by the bin duration, we obtain an estimate for the signal rate $r_s$. A threshold condition $r_s>r_{th}$ is then imposed to tag those bins that potentially contain a physical signal. In addition, we impose the requirement that the excess is separately observed by two or more detectors. The threshold rate $r_{th}$ is based on the trade-off of minimizing the number of false positives, whilst maximizing the sensitivity of the search and set equal to $r_{th}=30\ \mathrm{Hz}$, as motivated in Appendix \ref{sub:rate}. By imposing a fixed threshold rate, we account for the uncertainty on the characterization of the background rate described in Appendix \ref{sub:bg}. No additional criteria on the statistical excess of the rate are imposed, as the occurrence of a bin edge in the Bayesian block light curve already signifies that a statistically significant change in the rate has occurred.

We define an emission episode as a continuous period of increased emission in the background subtracted light curve. If a GRB has two or more emission episodes, we verify that the intermediate quiescent periods contain enough statistics to ensure that the rate has dropped back to the background level. Quiescent periods for which the Poisson uncertainty on the average background rate exceeds 5\% are disregarded. For a typical burst \footnote{Assuming a GRB for which two detectors contribute to the combined light curve and each detector has a background rate of $1000\ \mathrm{Hz}$.}, this corresponds to a lower limit on the allowed duration of the quiescent period of $\sim0.2\ \mathrm{s}$. Given the data, we can hence not constrain models predicting precursors less than a few times 0.1~s before the start of the prompt emission. Particularly in the case of short GRBs, such short delay precursors have been proposed due to e.g. tidal crust failure in binary neutron star mergers \cite{250,219,248}.

Having obtained our signal regions, we define the prompt signal phase as the emission episode with the largest photon fluence. If one or more emission episodes precede the prompt phase, we select them as GRB precursors. To verify that these early emission episodes are not caused by an unrelated weak transient at a different location, we compare the relative photon count ratios of the selected detectors. If the photons from the two emission episodes are coming from the same source and thus direction, their relative fraction should be the same between the different NaI detectors. In general, we find that the count ratios of precursors are consistent with those of the prompt emission, except for five potential false candidates with deviating values, presented in Appendix \ref{sub:count}. These five edge cases are included in our analysis, but marked as being potentially unphysical in the precursor catalog. As a final check in our selection procedure, the light curves of all GRBs for which precursor emission is found are inspected by eye. This allows to verify that the identified emission was based on a reliable fit to the background rate and that the GBM detector was in a stable operation mode at the time of the GRB.

\section{Results}
Applying our signal selection method on all 2684 bursts, we find 320 GRBs that were triggered by Fermi-GBM, but do not show a signal following our criteria. In the following, we therefore restrict ourselves to the 2364 bursts for which a signal is found.

Our analysis identified 244 precursor emission episodes spread over 217 GRBs. We thus find that 9\% of all GRBs have one or more precursors. Any given burst is observed to have at most 3 precursors. The number of bursts having 1, 2 and 3 precursors corresponds to 192, 23 and 2, respectively. Based on the number of signals observed in a background control region and considering the combined time preceding the prompt emission of all GRBs, equal to $2.1\cdot 10^6\ \mathrm{s}$, we estimate that the number of false positives in our analysis is 36.1$\pm$8.8, roughly 15\% of the full sample. A complete catalog containing the start time, the duration, and the time separation of the precursors with respect to the prompt phase is given in Appendix \ref{sub:cat} or can be accessed via the online tool \cite{GRBweb}.

\textit{Short GRBs.}
The selected GRBs can be subdivided based on the duration of the burst. While 14\% of the 2364 examined bursts are short GRBs, only four (1.8\%) of the 217 bursts with precursors are short GRBs \footnote{The precursor emission our analysis identified in the short burst bn130504314 was disregarded after a visual inspection of the light curve.}. For each of these 4 bursts, we observe that the precursors occur within 2~s before the prompt emission. All 4 short GRBs have a precursor that is shorter in duration than the prompt phase and their quiescent times are consistent with one another up to a factor $\sim3$. While limited in statistics, we note that the time intervals between the onset of the precursor and prompt emission are smaller than the $1.7\ \mathrm{s}$ time gap separating the gravitational waves and the gamma rays that were observed from GRB 170817A \cite{013}. These short time scales are consistent with the predictions of binary neutron star models \cite{142,219,248} and in line with the results from previous studies \cite{049,219}.

\begin{figure}[t]
\centering
{\includegraphics[width=0.95\linewidth]{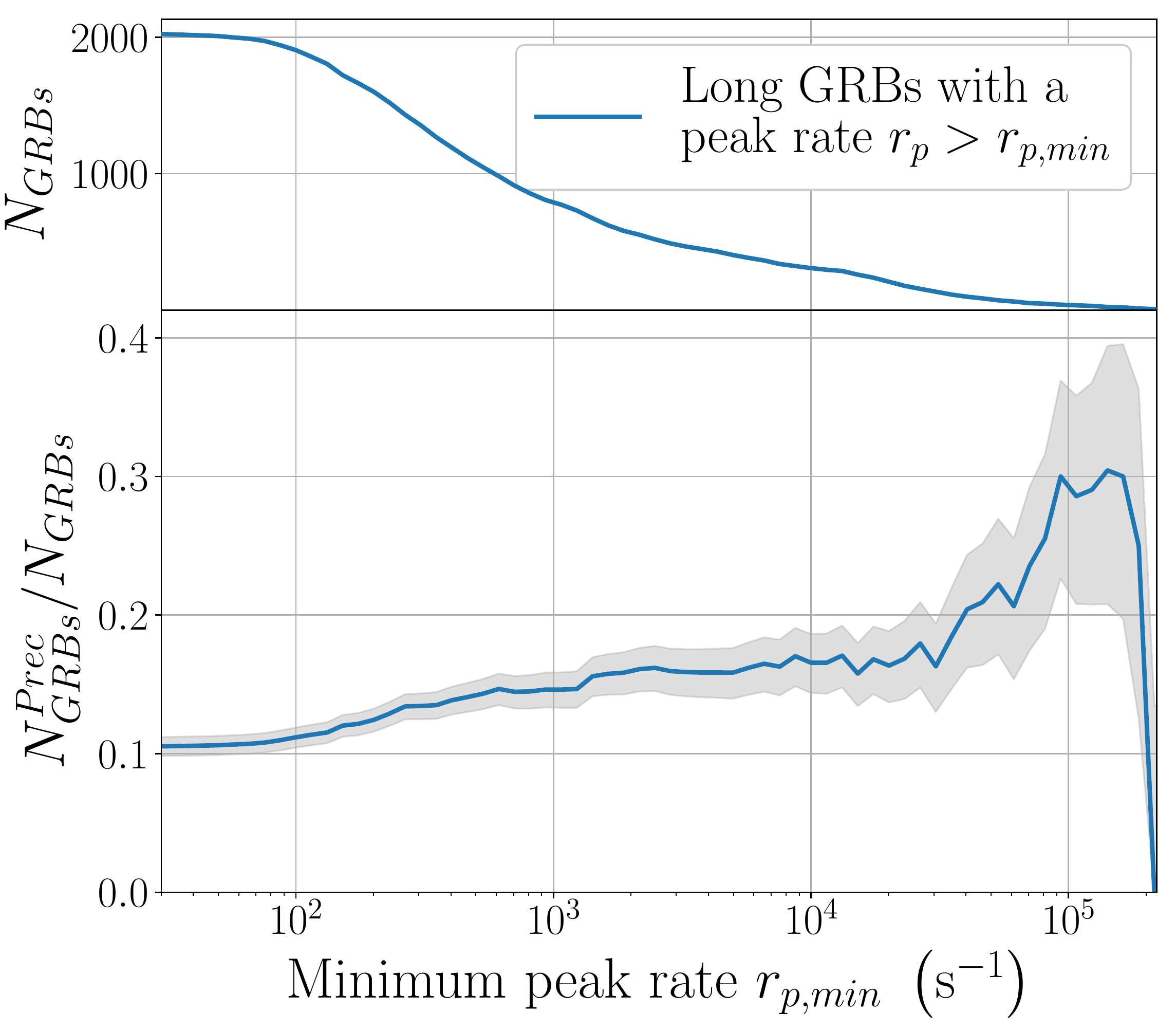}}
\caption{(top) Number of long GRBs whose peak rate exceeds the threshold value displayed on the $x$-axis. (bottom) Fraction of those GRBs for which precursors are observed. When selecting increasingly bright bursts, we observe an increase in the fraction of GRBs with precursors. The shaded grey band shows the 1$\sigma$ statistical uncertainty.}
\label{fig:PrecFracFluxCut}
\end{figure}

\textit{Long bright GRBs.}
An important parameter that affects whether or not a precursor is observed, is the apparent brightness of the burst. If a burst is too dim, the even dimmer precursor will be indistinguishable over the background. Alternatively, if the prompt emission of a dim burst has multiple peaks, the first peak(s) might mistakenly be identified as a precursor. These effects are eliminated in some analyses \cite{050,108,204,218}, by only selecting sufficiently bright bursts. To study the impact of including dim bursts in our sample, the bottom half of Fig.~\ref{fig:PrecFracFluxCut} shows the fraction of GRBs for which a precursor is observed, when only including bursts whose per detector peak rate exceeds the threshold rate $r_{p,min}$ displayed on the $x$-axis. Only long bursts are used, as the number of short GRBs with precursors is too limited to make a significant statement. An initial increase in the fraction of GRB precursors is observed for increasing threshold rates, which flattens out at 16\% once the threshold rate exceeds $2\cdot 10^3\ \mathrm{Hz}$. This effect indicates that the true fraction of long GRBs with precursors is underestimated when including dim bursts. At rates exceeding $3\cdot 10^4\ \mathrm{Hz}$, a second rise is observed. However, no significant statement can be made due to the limited number of increasingly bright bursts, as seen from the top half of Fig.~\ref{fig:PrecFracFluxCut}.

\begin{figure}[t]
\centering
{\includegraphics[width=0.95\linewidth]{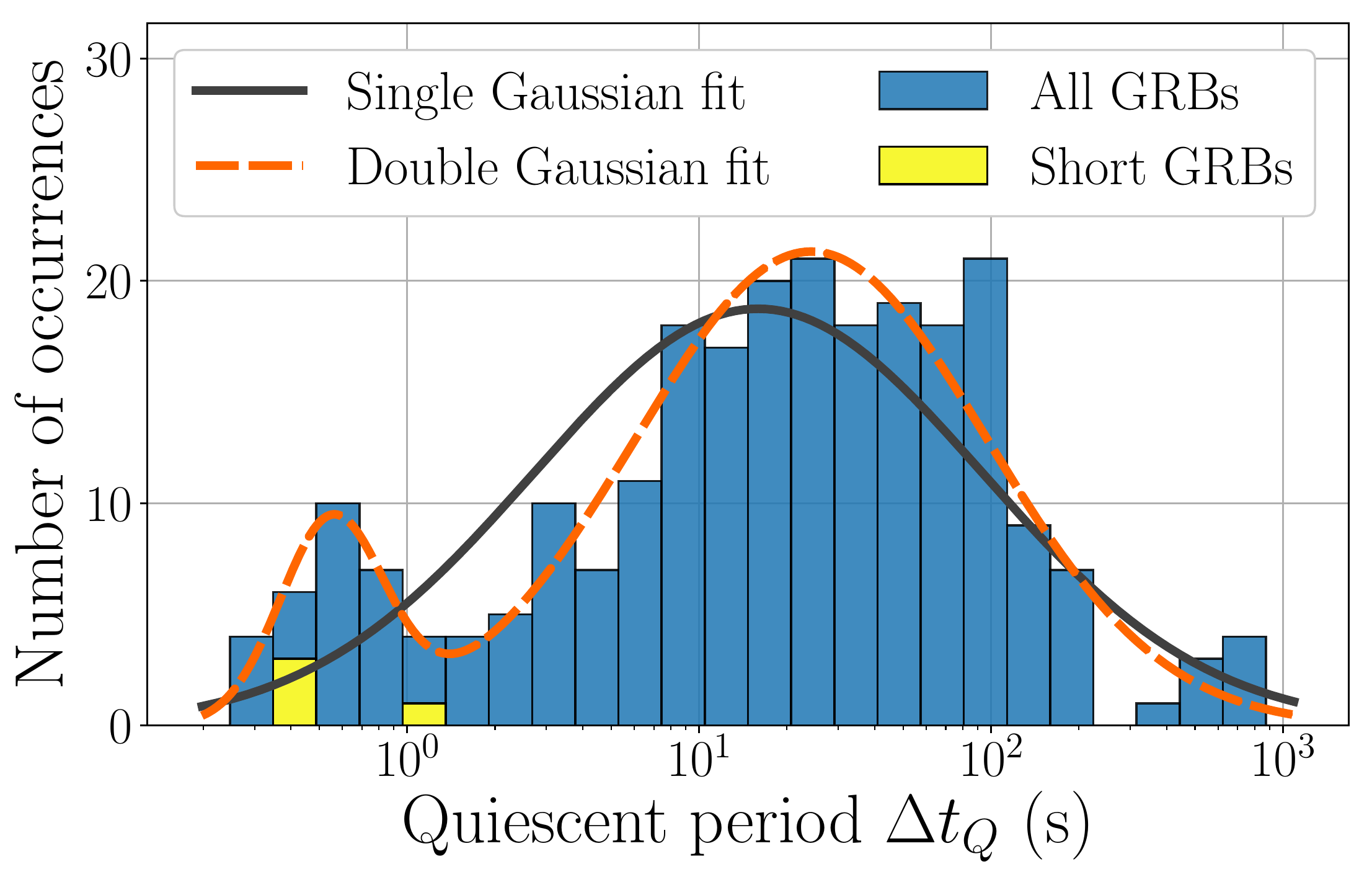}}
\caption{Distribution of the quiescent time between two subsequent emission episodes. The data is found to be well described by the sum of two Gaussian functions. The 4 short GRBs are indicated in yellow.}
\label{fig:Quiescent}
\end{figure}

\textit{Quiescent times.}
A quantity that can be used to relate the observed precursors to different theoretical models, is the duration of the quiescent interval separating two emission episodes. Figure~\ref{fig:Quiescent} displays the full distribution of the quiescent time that follows the observed precursors. Two populations are observed, crossing over at $\Delta t_Q\sim 1.4\ \mathrm{s}$. Applying a two-component Gaussian likelihood fit, we find that the distributions peak at $0.55\ \mathrm{s}$ and $24\ \mathrm{s}$ and have a weight of 11\% and 89\%, respectively. To evaluate the goodness-of-fit, we calculated the likelihood of obtaining the observed bin counts given the best fit parameters. Comparing this value to that of $10^6$ pseudo-experiments, randomly generated assuming Poisson statistics, we obtain a $p$-value $p=0.36$, indicating that a two-component Gaussian fit provides a good description for the data. Performing a single-component Gaussian fit, we obtain a $p$-value of only $p=8\cdot 10^{-5}$, showing the data is incompatible with a single Gaussian fit. An apparent third component shows up in the last three bins. This contribution is however most likely not physical, as the expected number of false positives is proportional to the width of the bins, thus linearly increasing from left to right.

\begin{figure}[t]
\centering
{\includegraphics[width=0.95\linewidth]{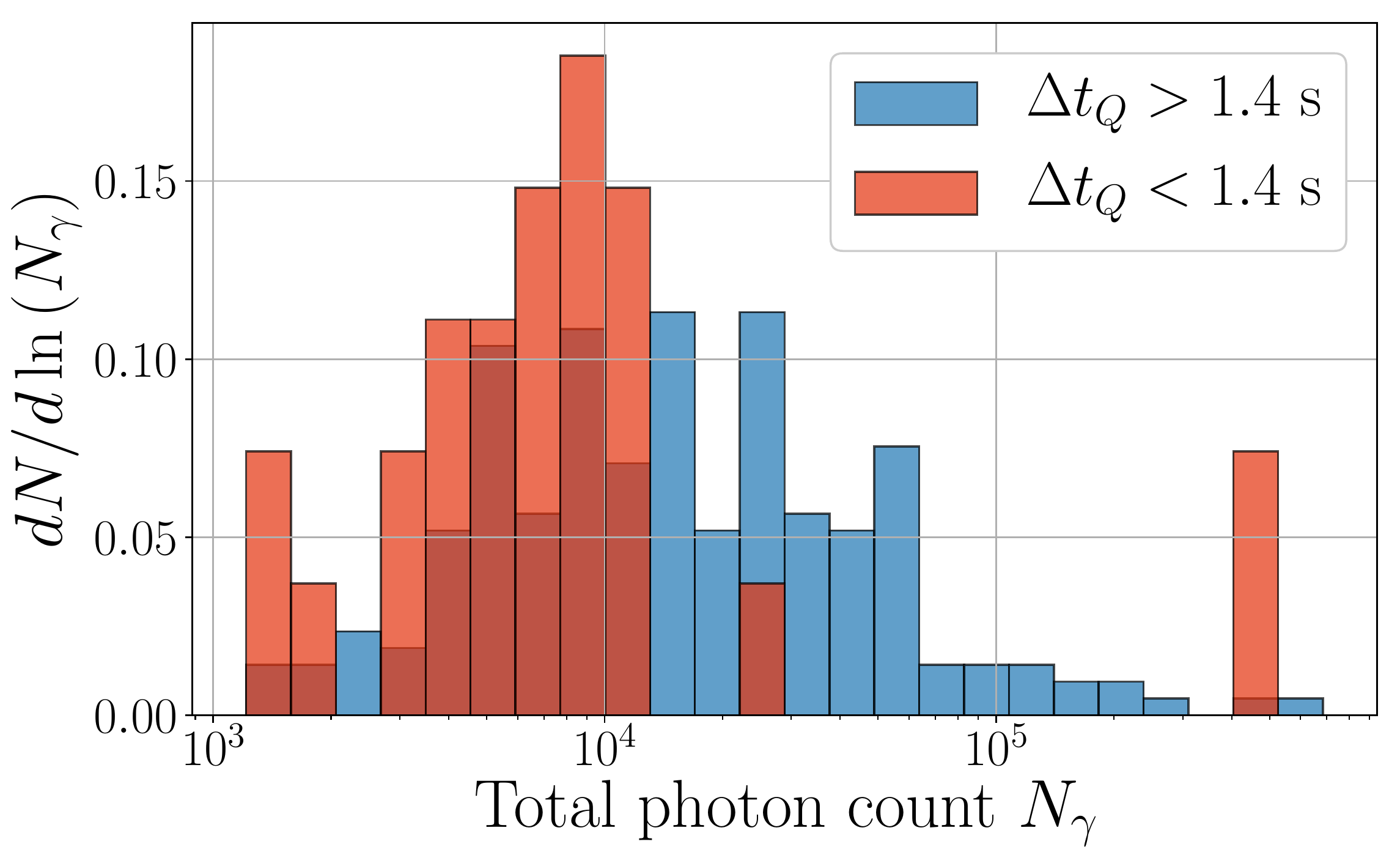}}
\caption{Probability distribution of the total photon count of long GRBs in which we identified precursors. The normalization is taken such that the y-values sum to unity. Bursts for which the quiescent times in between emission episodes is less than 1.4~s are observed to be less bright on average.}
\label{fig:DTQ_fluence}
\end{figure}

The leftmost component of the double Gaussian fit in Fig.~\ref{fig:Quiescent} could have several origins \footnote{None of the five potential false precursors with deviating count ratios contribute to the first component of the quiescent time distribution.}. A first contribution is found from the precursors of short GRBs, though they can only account for $\sim15\%$ of the observed excess. A second contribution could come from bursts whose observed flux drops below the observable limit in between different peaks of the prompt phase, thus falsely identifying as precursor emission. Figure~\ref{fig:DTQ_fluence} illustrates that bursts with $\Delta t_Q<1.4\ \mathrm{s}$ are on average less bright than bursts with longer quiescent periods. To probe the effect of dim bursts on the bi-modal feature, we repeated the analysis using only bright long GRBs, as detailed in Appendix \ref{sub:qui}. For this subset of bursts, we found that the two-component Gaussian fit ($p=0.57$) is still strongly preferred over the single Gaussian fit ($p=0.076$), although the obtained p-value for the single component Gaussian fit increased due to the decrease of statistics.

The observation of two components in the quiescent time distribution might indicate different physical origins for the short delay precursors given by the first component and the long delay precursors given by the second component of the fit. Short delay precursors to long GRBs could correspond to those bursts for which the photospheric emission is observed \cite{211}. In contrast, the longer quiescent times of tens to hundreds of seconds can be explained by models in which a jet is launched multiple times. The repeated launch of a jet has in some cases \cite{190} been confirmed by identifying the separate afterglow of the precursor and prompt emission. To uncover the origin of the short delay precursors, follow-up studies looking further into the individual properties of these events are encouraged.

\begin{figure}[t]
\centering
{\includegraphics[width=0.95\linewidth]{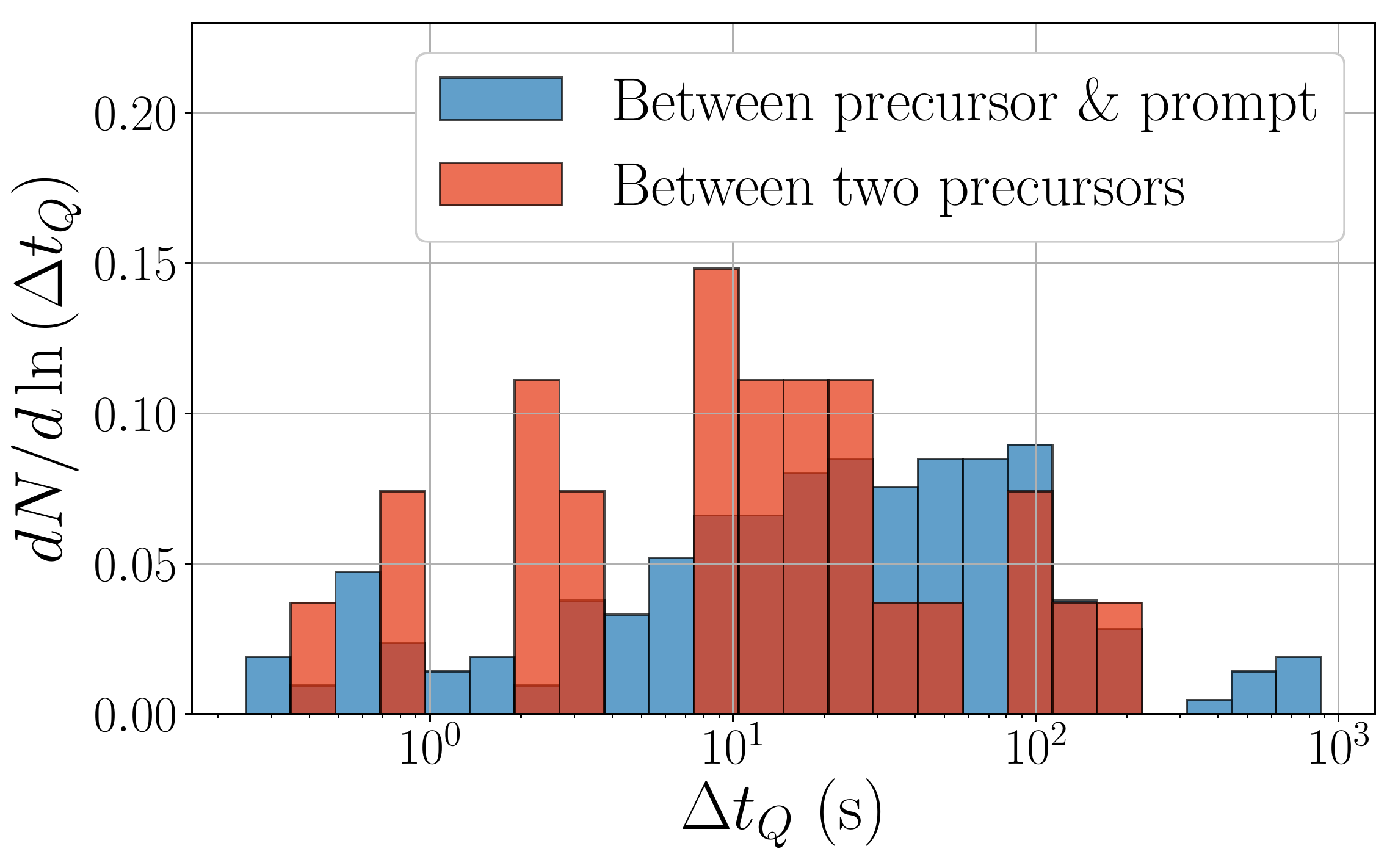}}
\caption{Probability distribution of the quiescent time $\Delta t_Q$ between two precursors (red) and between a precursor and the prompt emission (blue). The normalization is taken such that the y-values sum to unity.}
\label{fig:DTQ_prompt_vs_precursor}
\end{figure}

Quiescent times also provide an independent probe to investigate potential differences between the precursor and prompt emission. Previous studies generally found that precursor emission exhibits the same spectral properties as prompt emission \cite{049,110,107,108,162}. In the case of long GRBs, this observation can be embedded in a model in which the precursor and prompt emission arise from the same physical mechanism and are caused by the accretion of different shells of matter falling onto the central engine \cite{190,108,Zhang}. As such, there would be no intrinsic physical difference between precursor and prompt emission. Hence, the distribution of the quiescent times between two precursors and between precursor and prompt emission should be identical. These two distributions are shown in Fig.~\ref{fig:DTQ_prompt_vs_precursor}. To quantify their resemblance, we use a two-sample Kolmogorov-Smirnov test. The resulting $p$-value of $p=0.030$, while not significant, indicates that there potentially could be a difference between the two samples.

\textit{Temporal Correlations.}
A related study of quiescent times was performed in~\cite{204}, where a strong linear correlation between the duration of the quiescent time $\Delta t_Q$ and that of the subsequent emission episode $\Delta t_{sub}$ was found. This correlation was explained using a model in which potential energy builds up during the quiescent interval and is released once a critical threshold is reached. However, due to lack of data, redshift effects, which could naturally induce such a correlation, were not considered. To probe possible redshift effects, we  determined the correlation between $\Delta t_Q$ and $\Delta t_{sub}$ for the 21 bursts in our selection with known redshift $z$, and apply a correction for redshift. The obtained Pearson correlation factor is 34\%. To determine the significance of this value, we composed a test statistic distribution by calculating the correlation coefficient between random combinations of the quiescent times and secondary emission episodes. Based on this distribution, we obtain a $p$-value of $p=0.071$. No significant linear correlation is thus observed between the duration of the quiescent time following precursor episodes and the duration of the secondary emission episode. To test for a non-linear but monotonic correlation, we calculate the Spearman's rank coefficient $\rho$. Using the redshift corrected values of $\Delta t_Q$ and $\Delta t_{sub}$, we obtain $\rho=0.48$, corresponding to a $p$-value of $p=0.020$. A slight tension is thus observed for the null hypothesis that there is no correlation.

We applied the same methods to look for correlations between the duration of the precursor and prompt emission. As before, only bursts with known distance are used, such that redshift effects can be corrected for. A linear correlation coefficient of 21\% is recovered, corresponding to a $p$-value of $p=0.16$. Computing the Spearman's rank coefficient to probe for non-linear correlations, we obtain a correlation coefficient of 35\%, leading to a $p$-value of $p=0.11$. Hence, no significant correlation is observed.

\textit{GRB 190114C.}
One object in our selection is of special interest, GRB 190114C, a particularly bright burst that occurred on the 14th of January 2019 \cite{214}. GRB 190114C/bn190114873 is the first GRB from which TeV photons have been detected, as observed by the MAGIC telescope in La Palma \cite{194}. Our analysis identified two faint precursors occurring 5.57~s and 2.85~s before the start of the prompt emission and lasting 1.94~s and 1.54~s, respectively. The detailed light curve of this burst is shown in Fig.~\ref{fig:GRB190114C}.\newline

\section{Conclusion}
By applying a fully automated precursor search on the light curves of 2364 GRBs, we identified a total of 244 precursors spread over 217 bursts. Only four of those precursors occurred for short GRBs. We thus find that the fraction of long and short GRBs with one or more precursors equals 10.5\% and 1.2\%, respectively. All precursors for short GRBs occurred within $2\ \mathrm{s}$ before the start of the prompt emission. A notable long GRB for which we found two precursors is the extremely bright GRB 190114C. This burst was preceded by two dim precursors, indicating that gamma-ray production was already ongoing $5.6\  \mathrm{s}$ before the start of the prompt emission.

Apart from studying individual bursts, we also examined the quiescent time of all GRB precursors. A bi-modal distribution is observed, possibly indicating that precursors can have two types of progenitors. While we found no correlation between the duration of the precursor and prompt emission, we cannot exclude $(p=0.020)$ that there is a correlation between the quiescent time and the duration of the subsequent emission episode. Follow-up studies to further examine these claims are encouraged. To this end, and to allow other multi-messenger correlation studies, we have included a full list detailing the emission times of the identified precursors in Appendix \ref{sub:cat} and the online tool~\cite{GRBweb}. 

\begin{acknowledgments}
We are grateful for the public data made available by the Fermi-GBM collaboration. In addition, we would like to explicitly thank S. Zhu for providing a reference precursor catalog, the IceCube collaboration for hosting our online database, and the referees for their constructive feedback. This work was supported by the Flemish Foundation for Scientific Research (FWO-G007519N) and the European Research Council under the European Unions Horizon 2020 research and innovation program (No 805486 - K.~D. de Vries).
\end{acknowledgments}

\appendix
\section{Precursor catalog}\label{sub:cat}
To enable follow up studies, we provide a complete list of the emission times of all precursors identified by our analysis. Table \ref{tab:PrecursorTimes} provides the start time of the prompt emission in UTC, the start time of the precursor emission with respect to the onset of the prompt emission, and the duration of the precursor emission. An electronic version of this table can be downloaded from \url{https://icecube.wisc.edu/~grbweb_public/Precursors.html}.

\section{Background characterization}\label{sub:bg}
During normal operation, the Fermi telescope functions in a sky survey mode \cite{216}. This implies that the orientation of the spacecraft continuously changes to allow the LAT telescope to monitor the entire sky. A downside to this mode of operation is that the background rates of the GBM detectors are changing with time. A linear approximation can still be used over periods of time less than $\sim100\ \mathrm{s}$, as the period of the oscillatory motion of the spacecraft is on the order of 3 hours \cite{158,216}.

During previous searches, the time ranges used to estimate the background rate were generally set by hand \cite{107,050,110}. Since we plan to examine a time range of 2000~s for over 2000 bursts, this would become a very demanding endeavor. Therefore, we automated the selection of time intervals in which no increased gamma-ray activity is observed. This method has the added advantage that the selection is fully reproducible and based on physically motivated parameters.

\begin{figure*}[t]
{\includegraphics[width=\linewidth]{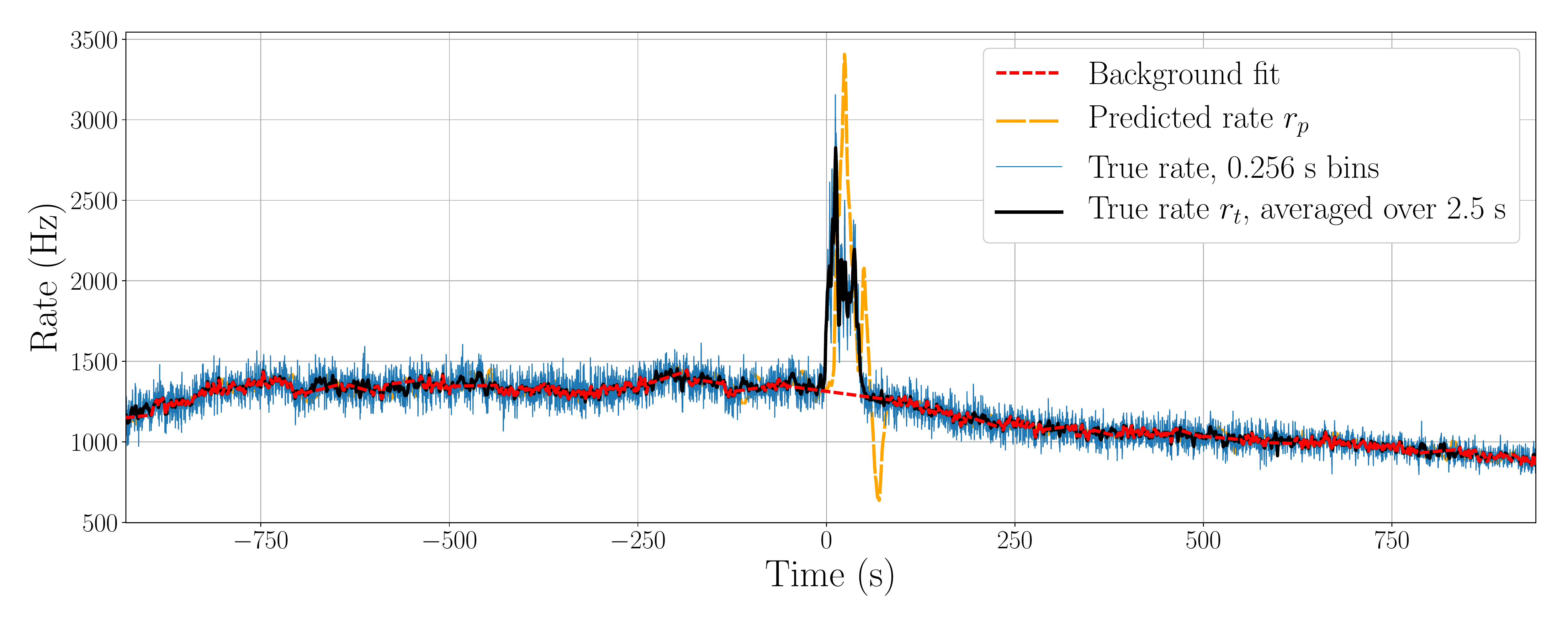}}
\caption{Characterization of the rate of GRB trigger bn150422703 and the GBM detector labeled nb. The true rate averaged over 2.5~s (black) is compared to a prediction based on prior data points (orange). Time intervals in which these two distributions match are used to characterize the background rate (red).}
\label{fig:Background}
\end{figure*}

Hence in this section, we will focus on the selection of good background intervals only. The tagging of reliable background time intervals is illustrated in Fig. \ref{fig:Background} and based on the assumption that the observed rate can be predicted using the rate at earlier times. To predict the background rate at an arbitrary time $t_1$, we perform a linear fit to the data in the time interval $[t_1-30\ \mathrm{s} , t_1-10\ \mathrm{s}]$. By extrapolating the fit to time $t_1$, we obtain a prediction $r_p$ for the background rate at time $t_1$. This prediction is then compared to the true rate $r_t$ found at time $t_1$, averaged over 2.5~s. As long as the true rate is within a 3$\sigma$ Poisson upper-fluctuation of the predicted background rate, i.e.
\begin{equation}
 r_t<r_p+3\cdot \sqrt{\frac{r_p}{2.5\ \mathrm{s}}}\ ,
 \label{eq:poisson}
\end{equation}
the time $t_1$ is tagged as background. The next point in time $t_2=(t_1+1\ \mathrm{s})$ is then subjected to the same procedure, until a time $t_n$ is found for which Eq.~\eqref{eq:poisson} no longer holds. Knowing that we have arrived at a possible non-background region, we immediately advance 25~s. This is done to overshoot the non-stable background period with possible excess emission. We then proceed by verifying if the RMS of $r_p-r_t$, averaged over a 10~s period centered around $t_n+25\ \mathrm{s}$ is within 1.5$\sigma$ of the Poisson expectation to verify background stability. If the RMS exceeds 1.5$\sigma$, $t_n+25\ \mathrm{s}$ is labeled as non-background. If, on the other hand, the RMS is sufficiently low, a new background interval is started at $t_n+25\ \mathrm{s}$. We then proceed with the procedure outlined above.

Using this method, background regions are identified in each of the light curves. The background rate is then set equal to the true rate, averaged over 2.5 s, in these background intervals. In intermediate possible regions of interest, a linear interpolation is used based on the last and first point of the adjacent background intervals. Figure~\ref{fig:Background} displays a visualization of this procedure for GRB trigger bn150422703.

\section{Poisson fluctuations}\label{sub:Poisson}
\begin{figure}[t]
\centering
{\includegraphics[width=\linewidth]{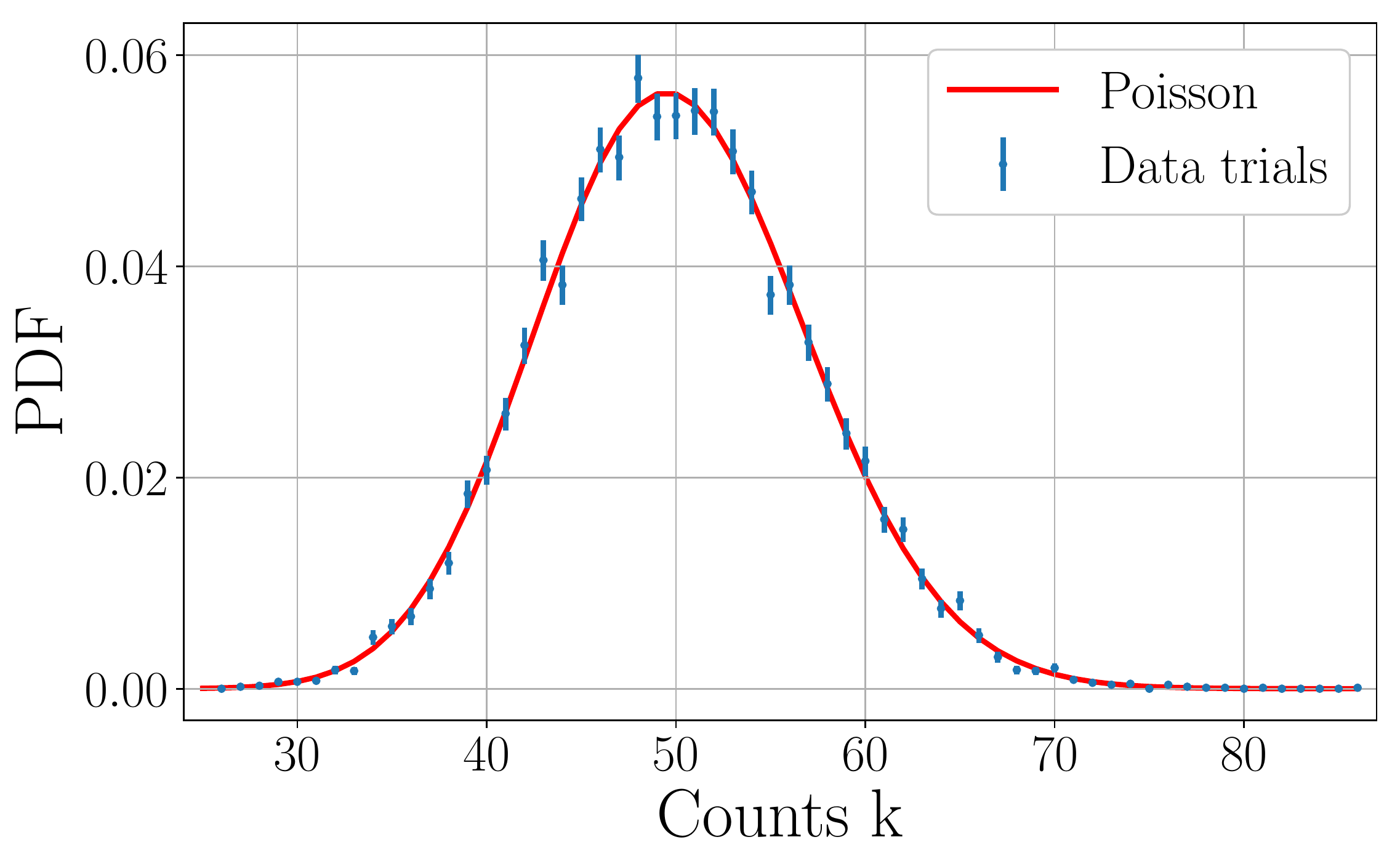}}
\caption{Probability distribution of the observed number of counts for 10.000 trials, compared to the theoretical Poisson distribution. The close agreement confirms that, on small time scales, the variation in the observed rate is Poissonian.}
\label{fig:Poisson_check_a}
\end{figure}

The method used to characterize the background rate builds on the assumption that the observed photon count follows Poisson statistics. To validate this assumption, we verify that the fluctuation in the number of observed photons describes a Poisson distribution. Using TTE data up to 30~s before the first signal is observed, we perform 10.000 trials, counting the number of photons observed during a time window $\Delta t=50\lambda$, typically $\sim$0.05~s. Here, $\lambda$ corresponds to the observed rate averaged over a 10~s period. Figure~\ref{fig:Poisson_check_a} displays the resulting distribution and the Poisson probability function expected from theory. The close match between the data and the theoretical expectation demonstrates that, on sufficiently small time scales, the variation in the observed photon count is Poissonian.

\begin{figure}[t]
\centering
{\includegraphics[width=\linewidth]{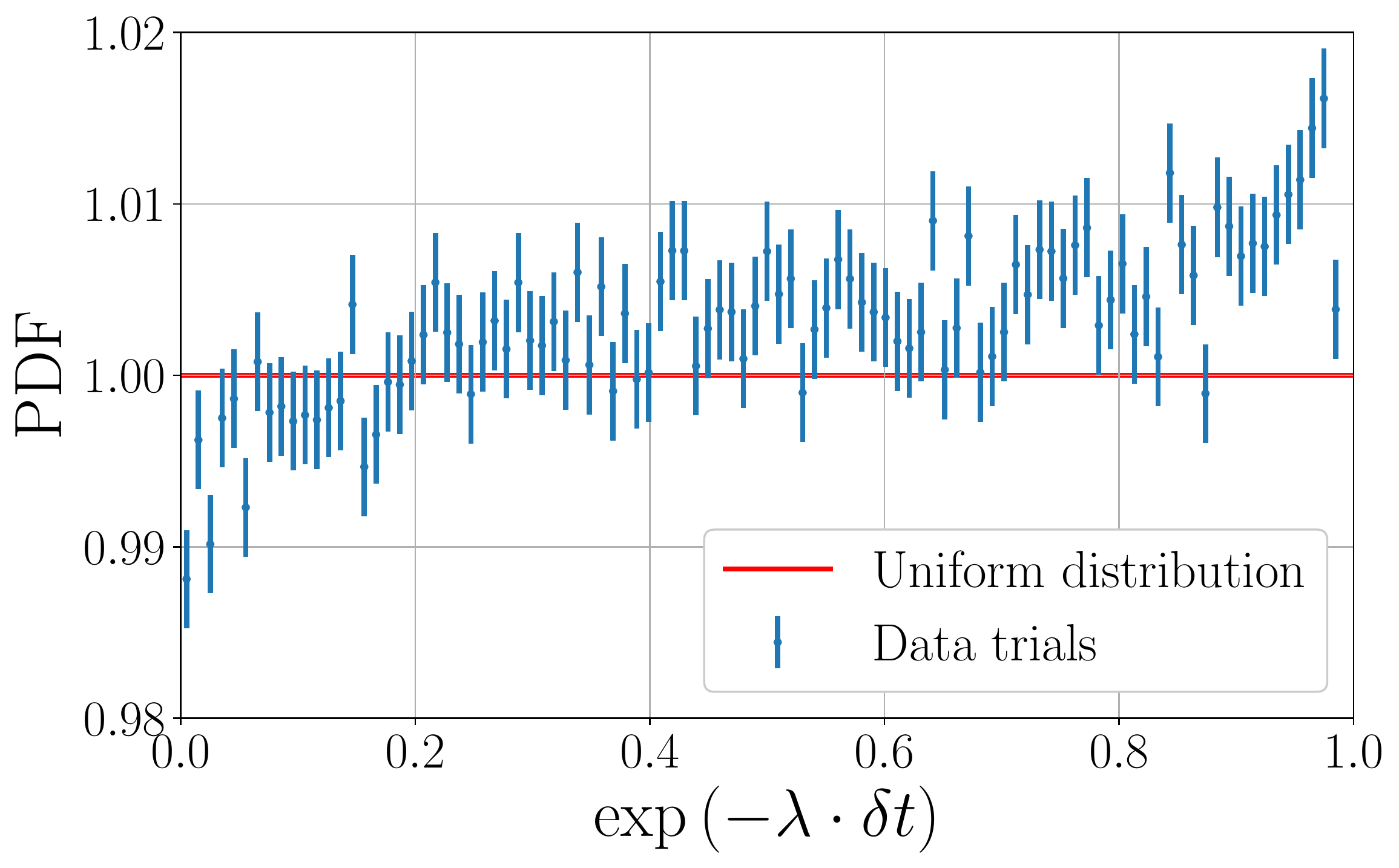}}
\caption{Probability distribution of $w=\exp(-\lambda\delta t)$, where $\delta t$ is the time delay between two subsequently observed photons and $\lambda$ is the average rate. The rightmost bin is not displayed as it has a $y$-value of only 0.7 due to detector dead time.}
\label{fig:Poisson_check_b}
\end{figure}

As a second check, we verify that the time delay $\delta t$ between two subsequently observed photons follows an exponential distribution. To this end, we apply the same procedure as before, taking 10.000 trials from background control regions with average rate $\lambda$. Each trial uses 1~s of data, meaning we obtain a total of $\sim 10^7$ values for $\delta t$. As $\delta t$ follows an exponential distribution, the variable $w=\exp(-\lambda\delta t)$ is uniformly distributed, where $w\in [0,1]$. Figure~\ref{fig:Poisson_check_b} shows the distribution of $w$, combining the data from all 10.000 trials. While some deviations are observed due to detector effects, the observed distribution matches the theoretical expectation within the 1\% level. The increase observed as $w\rightarrow 1$ is due to the pulse pile-up in the detector \cite{158}.

\section{Optimization of the threshold rate}\label{sub:rate}
To select bins that may contain a physical signal, we require that the background subtracted rate $r_s$ exceeds a threshold rate $r_{th}$. The Bayesian block algorithm already ensures that a statistically significant change of the rate is observed between adjacent bins. Hence, by imposing a fixed threshold rate, we mainly aim to account for the uncertainty that stems from the characterization of the background rate, as described in Appendix \ref{sub:bg}.

\begin{figure}[t]
\centering
{\includegraphics[width=\linewidth]{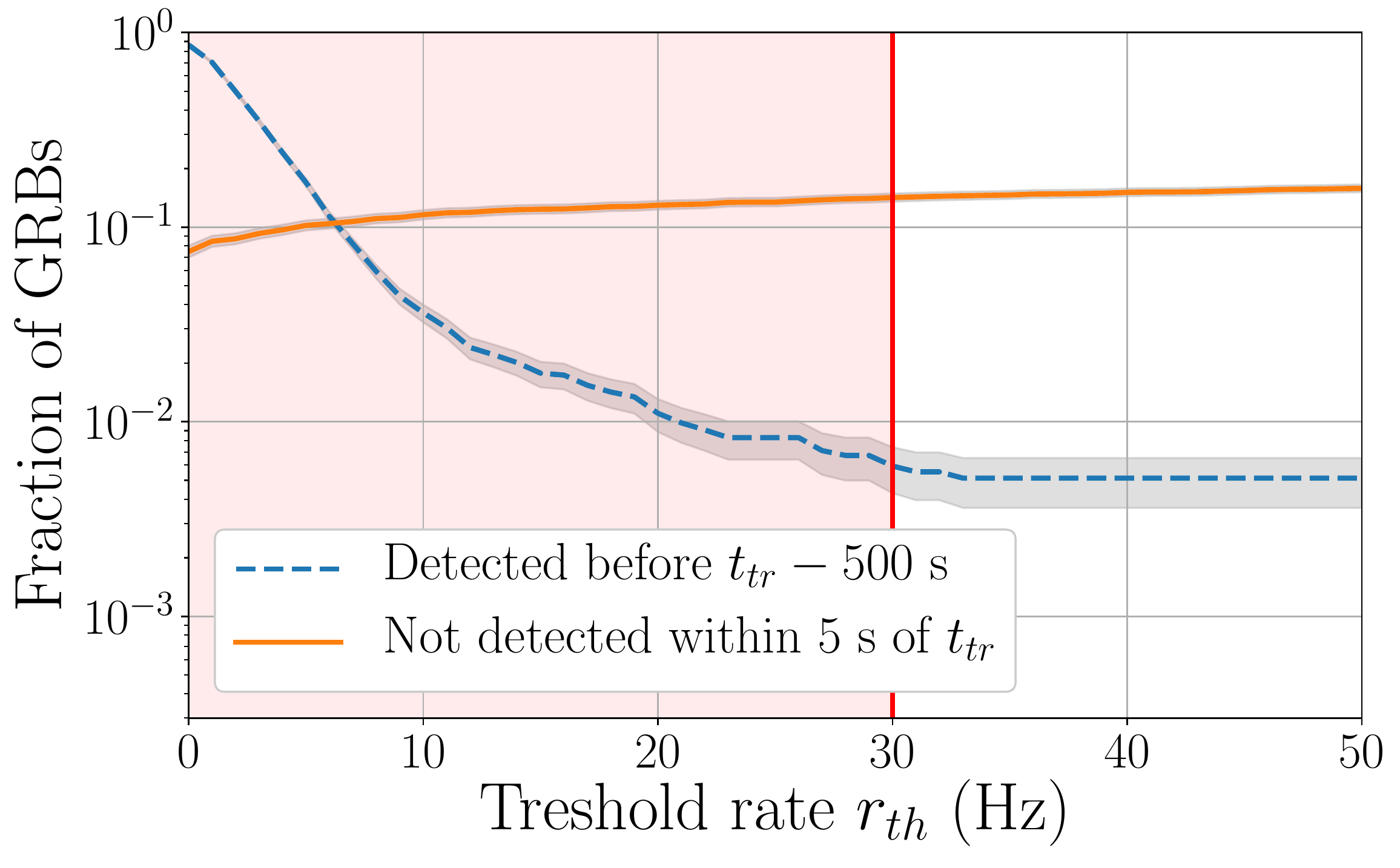}}
\caption{Relative number of GRBs that are not detected within 5~s of the trigger time $t_{tr}$ (full orange line) as a function of the threshold rate $r_{th}$. The same relation is shown for bursts that are detected more than 500~s before $t_{tr}$ (blue dashed line), where few to no precursors are expected \cite{109}. The shaded grey bands show the 1$\sigma$ statistical uncertainty for both curves. In our analysis, the threshold rate is set equal to $30\ \mathrm{Hz}$, indicated by the vertical line.}
\label{fig:RateSelection}
\end{figure}

The threshold rate $r_{th}$ is based on the trade-off of minimizing the number of false positives, whilst maximizing the sensitivity of the search. Ideally, every event triggering the GBM detector should also be selected by our analysis. To estimate the loss of sensitivity as a function of $r_{th}$, we therefore consider the fraction of GRBs in which, following our selection criteria, no excess is observed within 5~s of the GBM trigger time $t_{tr}$. This quantity is shown as the full orange line in Fig.~\ref{fig:RateSelection} and shows a slow but steady increase as a function of $r_{th}$.

To estimate the false positive rate, we consider the number of GRBs which, following our criteria, yielded an emission episode in the period from 1000~s to 500~s before the Fermi-GBM trigger time $t_{tr}$. This time window is based on a previous study presented in~\cite{109}, which, using a sample of 956 Fermi-GBM observed bursts, found only a single burst in which a precursor event occurred more than 500~s before $t_{tr}$. For our analysis, the fraction of GRBs which yielded a signal in this time range is displayed by the dashed blue line in Fig.~\ref{fig:RateSelection} as a function of the threshold rate $r_{th}$. A plateau is reached at $r_{th}>30\ \mathrm{Hz}$. We therefore set $r_{th}=30~\mathrm{Hz}$, as this corresponds to the minimal value for which the estimated false positive rate approaches the plateau at $\sim$0.5\%, leading to an expected false positive rate $r_f=1.7\cdot 10^{-5}\ \mathrm{Hz}$. On rare occasions, physical precursors have been observed more than 500~s before the prompt emission. One example is the case of GRB 091024 \cite{190}. As such, this approach is expected to result in a conservative estimate of the true false positive rate. 

\begin{figure}[t]
\centering
{\includegraphics[width=\linewidth]{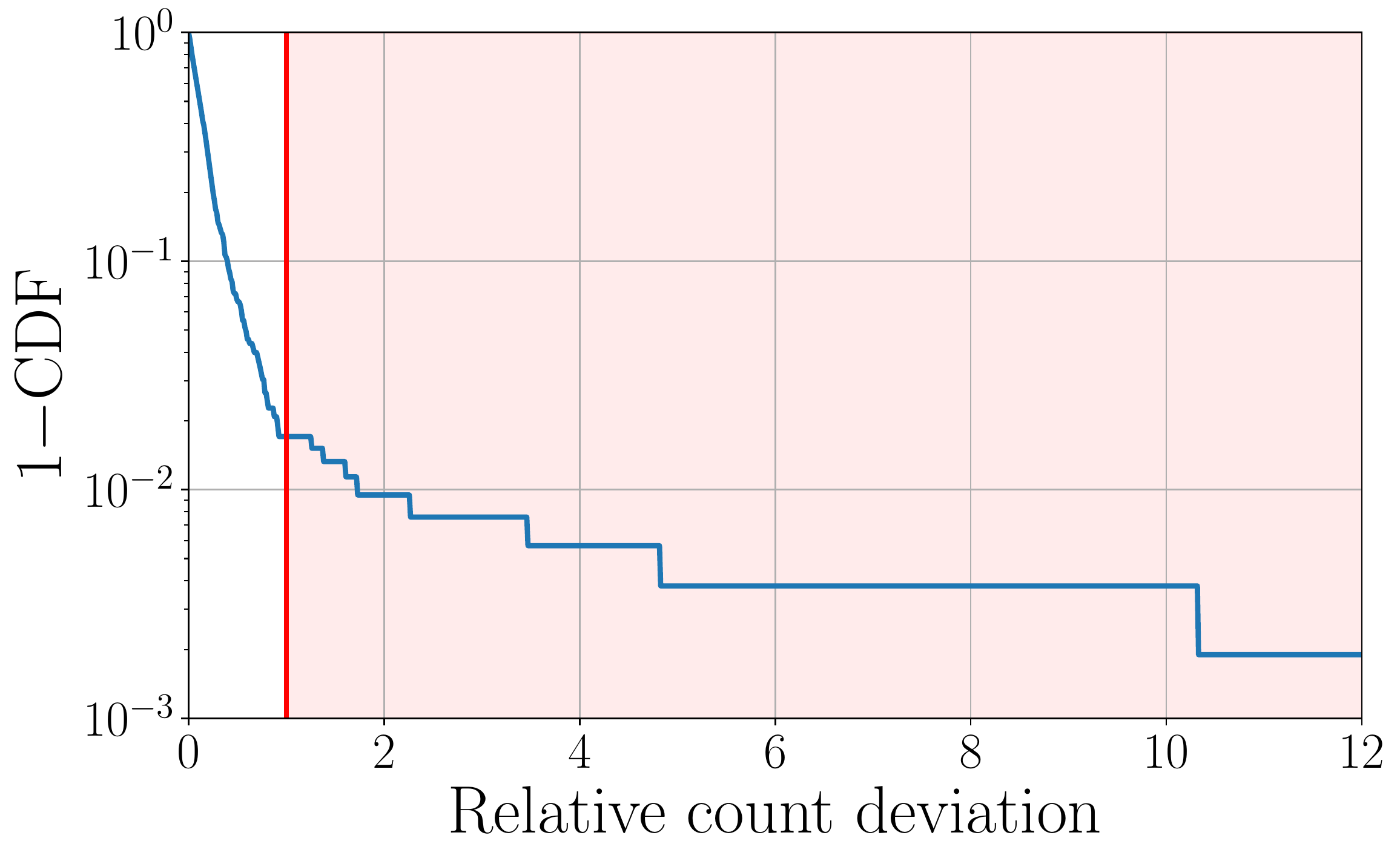}}
\caption{Distribution of the relative deviation of the count ratios, as defined in Equation \eqref{eq:CountRatios}. Five of the eight emission episodes that contribute to the extended tail cannot be confirmed to have a count ratio consistent with that of the prompt emission.}
\label{fig:CountRatios}
\end{figure}

\section{Relative count ratios}\label{sub:count}
Apart from physical GRB precursors, the possibility exist that an emission episode preceding the prompt phase is caused by an unrelated astrophysical transient. To test this hypothesis, we verify that the sky location of the two emission episodes are consistent with one another. Defining the relative count ratio $r$ to be the fraction of the total counts $N$ observed by a given detector $\alpha$, we compare the value of $r_\alpha$ between the precursor and prompt emission episode for each selected detector. Figure~\ref{fig:CountRatios} displays the combined distribution of
\begin{equation}
 \left| \frac{r_{\alpha,precursor}-r_{\alpha,prompt}}{r_{\alpha,prompt}} \right|
 \label{eq:CountRatios}
\end{equation}
for all identified precursors. An initial steep decline is observed, indicating good agreement between the count ratio of the precursor and prompt emission. An extended tail however shows up at relative deviations larger than one. For the eight bursts contributing to this tail, we inspected the light curves by eye. In the case of bn110428338, bn180307073 and bn180618724, the large excess could be resolved by improving the characterization of the background rate. This leaves five emission episodes, namely the precursor of bn090428441, bn110227229, bn130504314, bn150506398 and bn160908136, for which we cannot confirm that the location of the precursor is consistent with that of the prompt emission based on the relative count ratios.

\begin{figure}[t]
\centering
{\includegraphics[width=0.95\linewidth]{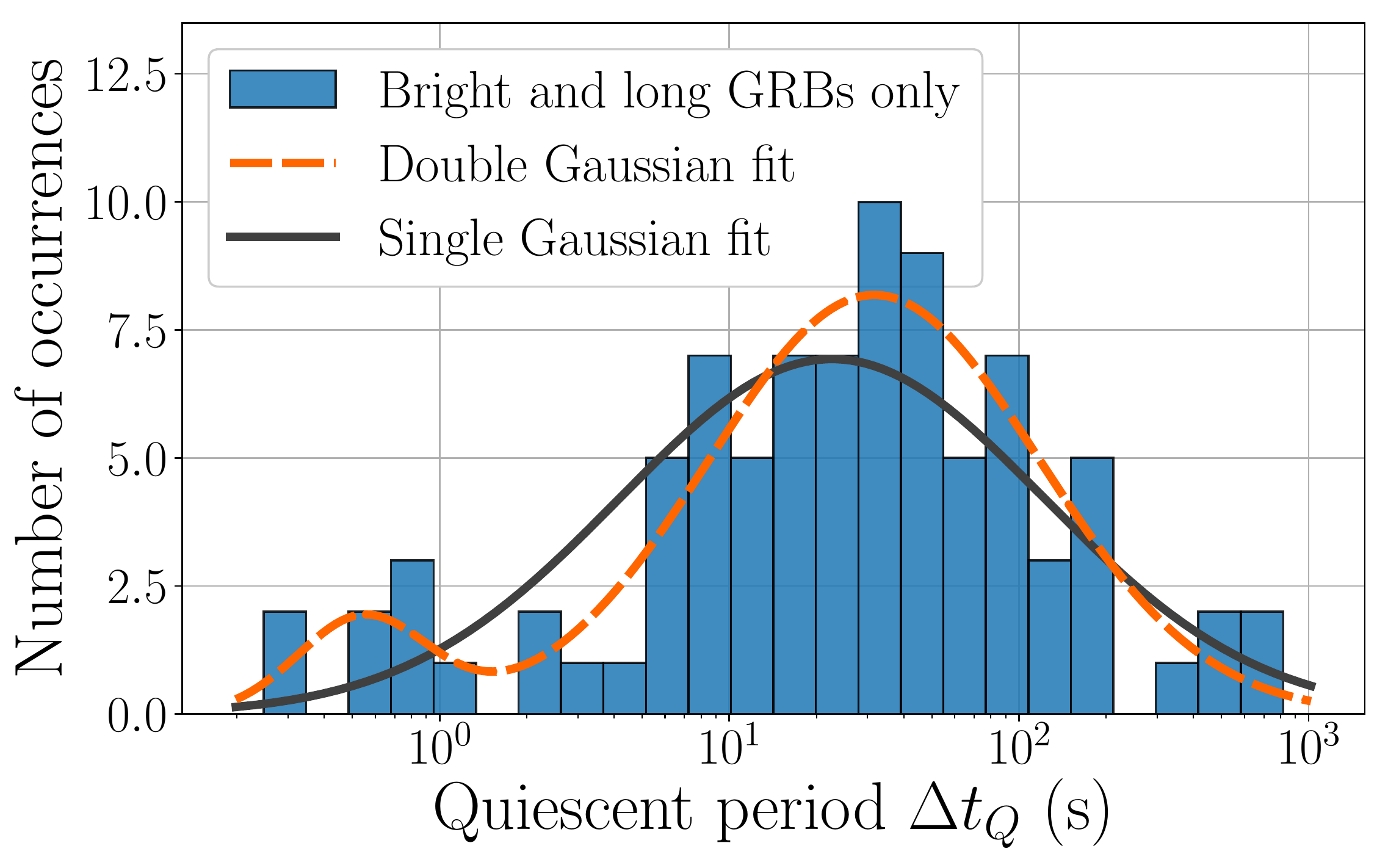}}
\caption{Distribution of the quiescent time between two subsequent emission episodes of bright and long GRBs. While statistics are significantly reduced, we find that a double component Gaussian fit is still preferred over a single Gaussian distribution.}
\label{fig:Quiescent2}
\end{figure}

\section{Quiescent time distribution}\label{sub:qui}
When examining the quiescent times between emission episodes, we found that the resulting distribution is well described by a two-component Gaussian fit. The inclusion of dim bursts in our sample could, however, lead to an artificial excess at short quiescent times, as the first peak of the prompt emission might be mistaken for a precursor. To probe this effect, we have repeated our analysis using only long bursts with a peak rate in excess of $3\cdot 10^3\ \mathrm{Hz}$. Figure~\ref{fig:Quiescent2} shows the resulting distribution. The two Gaussians now peak at $0.54\ \mathrm{s}$ and $32\ \mathrm{s}$ and have a weight of 8\% and 92\%, respectively. Based on the goodness-of-fit $p$-value, we find that the double component Gaussian distribution ($p=0.57$) is still preferred over the single-component Gaussian distribution ($p=0.076$). As expected, the disagreement between the data and single Gaussian fit is less significant than when using the full precursor sample due to the reduced sample size.

\onecolumngrid
\setlength{\LTleft}{-20cm plus -1fill}
\setlength{\LTright}{\LTleft}
\LTcapwidth=0.9\textwidth
\begin{longtable}{*{4}{p{4cm}}}
  \caption*{Table I: Temporal properties of the identified precursors. For every GRB, we provide the start time of the prompt emission $t_{prompt}$, the start time of the precursor emission with respect to $t_{prompt}$ and the duration of the precursor emission. The five potential false precursors with deviating count ratios are marked in \textit{italic}. To access this table in a digital format, please visit \url{https://icecube.wisc.edu/~grbweb_public/Precursors.html}.\label{tab:PrecursorTimes}} \\
  \toprule
  \bf{GRB} & \boldmath{$t_{prompt}$}\bf{ (UTC)} & \boldmath{$t_{precursor}$}\bf{ (s)} & \bf{Duration (s)} \\ \midrule
  \endfirsthead
  \caption*{Table \ref{tab:PrecursorTimes} continued: Temporal properties of the identified precursors.} \\
  \toprule
  \bf{GRB} & \boldmath{$t_{prompt}$}\bf{ (UTC)} & \boldmath{$t_{precursor}$}\bf{ (s)} & \bf{Duration (s)} \\ \midrule
  \endhead
  \bottomrule \multicolumn{4}{r}{\textit{Continued on next page}} \\
  \endfoot
  \bottomrule
  \endlastfoot
  \obeylines   bn080723557 & 13:22:55.412 & -34.284 & 28.319 \\ 
  bn080807993 & 23:50:44.177 & -11.612 & 1.032 \\ 
  bn080816503 & 12:04:39.495 & -21.823 & 1.823 \\ 
  bn080818579 & 13:54:44.589 & -20.361 & 5.596 \\ 
  bn080830368 & 08:50:22.699 & -8.559 & 5.112 \\ 
  bn081003644 & 15:27:27.738 & -11.363 & 4.320 \\ 
  bn081121858 & 20:35:31.671 & -8.498 & 7.855 \\ 
  bn090101758 & 18:13:07.574 & -86.950 & 6.082 \\ 
  bn090113778 & 18:40:38.870 & -0.475 & 0.150 \\ 
  bn090117335 & 08:02:26.183 & -24.653 & 1.296 \\ 
  bn090131090 & 02:09:43.196 & -22.324 & 12.445 \\ 
  bn090309767 & 18:25:41.699 & -36.134 & 6.122 \\ 
  bn090326633 & 15:10:16.566 & -583.057 & 0.256 \\ 
  bn090326633 & 15:10:16.566 & -580.753 & 5.376 \\ 
  bn090419997 & 23:55:38.751 & -37.348 & 23.251 \\ 
  bn090425377 & 09:04:14.740 & -44.805 & 2.705 \\ 
  \textit{bn090428441} & 10:34:37.862 & -26.762 & 18.048 \\ 
  bn090502777 & 18:40:11.917 & -37.539 & 3.065 \\ 
  bn090510016 & 00:23:00.368 & -0.420 & 0.024 \\ 
  bn090602564 & 13:32:22.296 & -1.242 & 0.683 \\ 
  bn090610723 & 17:22:58.385 & -90.937 & 6.686 \\ 
  bn090618353 & 08:29:16.651 & -50.628 & 28.946 \\ 
  bn090720710 & 17:02:57.665 & -0.776 & 0.264 \\ 
  bn090810659 & 15:50:40.542 & -94.594 & 43.262 \\ 
  bn090811696 & 16:41:54.351 & -4.958 & 1.583 \\ 
  bn090814950 & 22:48:30.233 & -43.778 & 18.577 \\ 
  bn090815946 & 22:44:41.956 & -179.466 & 12.722 \\ 
  bn090820509 & 12:13:25.368 & -8.951 & 4.124 \\ 
  bn090907017 & 00:24:10.767 & -1.967 & 1.664 \\ 
  bn090929190 & 04:33:04.488 & -0.571 & 0.122 \\ 
  bn091109895 & 21:28:49.421 & -9.606 & 2.788 \\ 
  bn100116897 & 21:32:19.006 & -83.382 & 6.319 \\ 
  bn100130729 & 17:30:19.867 & -65.378 & 23.215 \\ 
  bn100204566 & 13:34:36.243 & -16.948 & 15.677 \\ 
  bn100323542 & 13:01:32.005 & -54.935 & 9.109 \\ 
  bn100326402 & 09:37:30.596 & -55.808 & 32.512 \\ 
  bn100424876 & 21:03:54.875 & -123.791 & 2.521 \\ 
  bn100517154 & 03:42:30.304 & -22.365 & 1.362 \\ 
  bn100619015 & 00:22:24.001 & -77.870 & 9.918 \\ 
  bn100625891 & 21:22:58.362 & -15.645 & 4.029 \\ 
  bn100709602 & 14:28:25.731 & -56.254 & 16.328 \\ 
  bn100718160 & 03:50:13.287 & -25.036 & 6.090 \\ 
  bn100718160 & 03:50:13.287 & -7.415 & 6.808 \\ 
  bn100730463 & 11:06:50.220 & -41.808 & 12.805 \\ 
  bn100730463 & 11:06:50.220 & -18.243 & 0.001 \\ 
  bn100827455 & 10:55:49.710 & -0.442 & 0.079 \\ 
  bn100923844 & 20:15:31.462 & -24.128 & 4.019 \\ 
  bn101030664 & 15:56:24.411 & -69.697 & 31.744 \\ 
  bn101224578 & 13:53:30.861 & -33.455 & 10.658 \\ 
  bn101227536 & 12:51:49.785 & -3.895 & 3.646 \\ 
  bn110102788 & 18:55:41.740 & -67.434 & 25.256 \\ 
  \textit{bn110227229} & 05:30:09.611 & -111.145 & 21.120 \\ 
  bn110428338 & 08:07:18.821 & -70.448 & 42.874 \\ 
  bn110428338 & 08:07:18.821 & -18.748 & 13.398 \\ 
  bn110528624 & 14:59:12.297 & -217.477 & 11.264 \\ 
  bn110528624 & 14:59:12.297 & -35.653 & 13.654 \\ 
  bn110528624 & 14:59:12.297 & -21.303 & 13.839 \\ 
  bn110725236 & 05:39:57.932 & -16.720 & 7.619 \\ 
  bn110729142 & 03:30:47.288 & -342.504 & 52.731 \\ 
  bn110729142 & 03:30:47.288 & -185.188 & 51.556 \\ 
  bn110825102 & 02:26:58.702 & -7.864 & 0.814 \\ 
  bn110903111 & 02:42:41.553 & -187.466 & 22.062 \\ 
  bn110904124 & 02:58:55.085 & -44.632 & 7.665 \\ 
  bn110909116 & 02:47:01.914 & -4.433 & 1.670 \\ 
  bn110926107 & 02:34:30.183 & -45.717 & 3.110 \\ 
  bn111010709 & 17:01:07.319 & -34.749 & 31.018 \\ 
  bn111015427 & 10:15:22.011 & -25.770 & 17.144 \\ 
  bn111228657 & 15:45:16.506 & -46.111 & 10.496 \\ 
  bn111228657 & 15:45:16.506 & -32.543 & 11.776 \\ 
  bn111230683 & 16:23:06.415 & -11.301 & 4.631 \\ 
  bn111230819 & 19:39:41.521 & -9.814 & 1.304 \\ 
  bn111230819 & 19:39:41.521 & -8.120 & 4.234 \\ 
  bn120118709 & 17:00:24.779 & -6.498 & 5.475 \\ 
  bn120308588 & 14:06:05.511 & -21.363 & 3.092 \\ 
  bn120319983 & 23:35:18.709 & -17.629 & 5.551 \\ 
  bn120412920 & 22:05:51.344 & -71.057 & 5.502 \\ 
  bn120504945 & 22:40:07.713 & -1.369 & 0.799 \\ 
  bn120513531 & 12:44:14.932 & -15.008 & 1.330 \\ 
  bn120530121 & 02:54:31.969 & -50.475 & 7.974 \\ 
  bn120611108 & 02:35:54.181 & -8.321 & 6.602 \\ 
  bn120710100 & 02:25:09.865 & -113.086 & 4.857 \\ 
  bn120711115 & 02:45:52.633 & -61.735 & 4.838 \\ 
  bn120716712 & 17:08:00.170 & -176.365 & 5.383 \\ 
  bn120819048 & 01:09:20.076 & -60.316 & 7.618 \\ 
  bn120819048 & 01:09:20.076 & -30.405 & 1.638 \\ 
  bn121005340 & 08:10:54.001 & -101.730 & 38.794 \\ 
  bn121029350 & 08:24:27.774 & -11.090 & 8.798 \\ 
  bn121031949 & 22:50:21.029 & -191.769 & 38.495 \\ 
  bn121113544 & 13:03:25.589 & -45.362 & 31.652 \\ 
  bn121125356 & 08:32:50.026 & -29.374 & 20.325 \\ 
  bn121217313 & 07:29:53.089 & -714.103 & 65.792 \\ 
  bn130104721 & 17:18:12.706 & -5.969 & 3.898 \\ 
  bn130106995 & 23:52:56.117 & -33.325 & 17.558 \\ 
  bn130208684 & 16:24:43.858 & -21.975 & 5.099 \\ 
  bn130209961 & 23:03:46.502 & -5.102 & 4.597 \\ 
  bn130219775 & 18:36:47.745 & -56.310 & 20.260 \\ 
  bn130310840 & 20:09:45.591 & -4.755 & 1.194 \\ 
  bn130318456 & 10:57:50.305 & -82.735 & 6.897 \\ 
  bn130320560 & 13:29:06.051 & -159.315 & 42.085 \\ 
  bn130404840 & 20:10:25.030 & -21.354 & 8.355 \\ 
  bn130418844 & 20:16:08.506 & -87.313 & 16.452 \\ 
  \textit{bn130504314} & 07:32:36.037 & -32.672 & 0.464 \\ 
  bn130623130 & 03:07:03.470 & -26.744 & 1.821 \\ 
  bn130720582 & 13:59:14.940 & -146.139 & 115.366 \\ 
  bn130813791 & 18:59:18.842 & -5.810 & 1.680 \\ 
  bn130815660 & 15:51:22.993 & -31.482 & 6.925 \\ 
  bn130818941 & 22:34:29.441 & -70.463 & 8.706 \\ 
  bn130919173 & 04:09:40.924 & -0.686 & 0.236 \\ 
  bn131014513 & 12:18:34.911 & -20.917 & 2.089 \\ 
  bn131108024 & 00:34:43.981 & -2.395 & 1.815 \\ 
  bn140104731 & 17:34:01.991 & -120.439 & 66.204 \\ 
  bn140104731 & 17:34:01.991 & -24.501 & 1.459 \\ 
  bn140108721 & 17:19:53.720 & -71.900 & 11.570 \\ 
  bn140126815 & 19:33:40.215 & -62.234 & 20.478 \\ 
  bn140126815 & 19:33:40.215 & -24.368 & 14.110 \\ 
  bn140304849 & 20:25:37.760 & -189.609 & 30.654 \\ 
  bn140329295 & 07:04:57.833 & -19.534 & 0.630 \\ 
  bn140404030 & 00:43:22.825 & -71.917 & 7.657 \\ 
  bn140512814 & 19:33:23.687 & -98.421 & 11.788 \\ 
  bn140621827 & 19:50:14.988 & -4.111 & 0.718 \\ 
  bn140628704 & 16:54:21.456 & -66.005 & 4.910 \\ 
  bn140709051 & 01:13:51.906 & -11.597 & 5.700 \\ 
  bn140714268 & 06:27:35.035 & -109.468 & 27.544 \\ 
  bn140716436 & 10:29:26.513 & -89.084 & 2.218 \\ 
  bn140818229 & 05:31:17.613 & -69.604 & 10.233 \\ 
  bn140824606 & 14:34:24.964 & -73.928 & 12.933 \\ 
  bn140825328 & 07:53:42.446 & -59.289 & 11.821 \\ 
  bn140825328 & 07:53:42.446 & -38.258 & 3.215 \\ 
  bn140917512 & 12:17:10.292 & -4.434 & 3.940 \\ 
  bn141029134 & 03:14:24.675 & -66.449 & 3.739 \\ 
  bn141029134 & 03:14:24.675 & -41.574 & 6.940 \\ 
  bn141102536 & 12:51:40.471 & -1.269 & 0.088 \\ 
  bn150126868 & 20:51:32.131 & -55.037 & 13.019 \\ 
  bn150127398 & 09:32:49.909 & -6.512 & 5.747 \\ 
  bn150226545 & 13:08:44.224 & -202.152 & 1.028 \\ 
  bn150226545 & 13:08:44.224 & -155.467 & 7.878 \\ 
  bn150226545 & 13:08:44.224 & -41.188 & 16.158 \\ 
  bn150330828 & 19:53:59.254 & -98.194 & 11.512 \\ 
  bn150416773 & 18:33:22.811 & -824.534 & 42.496 \\ 
  bn150422703 & 16:52:31.997 & -468.581 & 15.616 \\ 
  \textit{bn150506398} & 09:33:46.679 & -116.285 & 27.791 \\ 
  bn150508945 & 22:40:36.620 & -102.265 & 15.712 \\ 
  bn150512432 & 10:23:46.759 & -86.467 & 43.029 \\ 
  bn150512432 & 10:23:46.759 & -28.593 & 20.212 \\ 
  bn150522433 & 10:24:07.264 & -19.511 & 7.822 \\ 
  bn150523396 & 09:30:14.993 & -28.370 & 19.748 \\ 
  bn150627183 & 04:23:22.017 & -458.665 & 3.072 \\ 
  bn150702998 & 23:56:45.108 & -6.691 & 2.490 \\ 
  bn150703149 & 03:33:54.082 & -13.280 & 0.008 \\ 
  bn150830128 & 03:04:38.646 & -14.638 & 14.021 \\ 
  bn151027166 & 04:00:00.254 & -96.360 & 40.571 \\ 
  bn151030999 & 23:59:47.634 & -88.314 & 17.686 \\ 
  bn151211672 & 16:07:28.188 & -151.405 & 26.022 \\ 
  bn160131174 & 04:12:52.609 & -179.691 & 44.007 \\ 
  bn160201883 & 21:11:44.177 & -1.590 & 0.968 \\ 
  bn160215773 & 18:36:08.605 & -109.239 & 44.645 \\ 
  bn160219673 & 16:11:34.712 & -110.393 & 12.546 \\ 
  bn160223072 & 01:45:54.364 & -95.615 & 10.496 \\ 
  bn160225809 & 19:25:09.731 & -48.115 & 23.215 \\ 
  bn160512199 & 04:45:57.662 & -56.663 & 9.377 \\ 
  bn160519012 & 00:18:55.054 & -83.260 & 3.345 \\ 
  bn160519012 & 00:18:55.054 & -65.164 & 17.101 \\ 
  bn160523919 & 22:04:13.977 & -38.410 & 5.424 \\ 
  bn160625945 & 22:43:14.090 & -178.317 & 2.418 \\ 
  bn160724444 & 10:40:02.521 & -7.324 & 1.790 \\ 
  bn160821857 & 20:36:22.642 & -117.067 & 31.832 \\ 
  bn160825799 & 19:10:50.313 & -1.449 & 0.599 \\ 
  \textit{bn160908136} & 03:16:48.679 & -87.733 & 6.845 \\ 
  bn160912521 & 12:31:42.840 & -57.422 & 36.635 \\ 
  bn160912521 & 12:31:42.840 & -17.072 & 5.193 \\ 
  bn160919613 & 14:43:36.685 & -24.729 & 0.498 \\ 
  bn160919613 & 14:43:36.685 & -15.527 & 0.761 \\ 
  bn161105417 & 10:01:18.575 & -30.217 & 12.749 \\ 
  bn161111197 & 04:44:50.633 & -102.555 & 11.125 \\ 
  bn161117066 & 01:37:14.177 & -103.474 & 77.027 \\ 
  bn161119633 & 15:11:02.131 & -10.916 & 7.666 \\ 
  bn161129300 & 07:11:45.292 & -5.373 & 0.040 \\ 
  bn170109137 & 03:21:41.186 & -245.940 & 18.163 \\ 
  bn170109137 & 03:21:41.186 & -217.040 & 6.377 \\ 
  bn170115662 & 15:54:01.580 & -95.287 & 18.563 \\ 
  bn170209048 & 01:09:05.007 & -28.188 & 8.228 \\ 
  bn170302719 & 17:15:41.992 & -22.294 & 12.259 \\ 
  bn170323775 & 18:36:31.186 & -12.963 & 12.697 \\ 
  bn170402961 & 23:03:40.777 & -15.936 & 1.501 \\ 
  bn170402961 & 23:03:40.777 & -12.442 & 0.230 \\ 
  bn170416583 & 14:00:34.758 & -35.298 & 12.494 \\ 
  bn170514152 & 03:38:43.989 & -5.895 & 0.678 \\ 
  bn170514180 & 04:19:54.177 & -79.666 & 35.908 \\ 
  bn170830069 & 01:38:59.546 & -19.395 & 5.987 \\ 
  bn170831179 & 04:18:03.061 & -73.621 & 8.547 \\ 
  bn170831179 & 04:18:03.061 & -43.400 & 6.309 \\ 
  bn170923188 & 04:31:15.015 & -10.012 & 1.018 \\ 
  bn171004857 & 20:33:34.433 & -2.263 & 1.378 \\ 
  bn171102107 & 02:34:03.231 & -29.516 & 10.393 \\ 
  bn171112868 & 20:50:13.004 & -198.952 & 8.192 \\ 
  bn171112868 & 20:50:13.004 & -43.928 & 9.502 \\ 
  bn171120556 & 13:20:33.596 & -31.460 & 4.221 \\ 
  bn171211844 & 20:17:18.932 & -82.541 & 12.393 \\ 
  bn180124392 & 09:23:59.613 & -4.987 & 0.611 \\ 
  bn180126095 & 02:16:29.991 & -820.685 & 11.776 \\ 
  bn180307073 & 01:44:35.183 & -39.275 & 23.342 \\ 
  bn180411519 & 12:28:28.650 & -54.086 & 26.673 \\ 
  bn180416340 & 08:10:01.701 & -36.541 & 10.291 \\ 
  bn180426549 & 13:10:59.907 & -13.182 & 5.544 \\ 
  bn180618724 & 17:22:55.701 & -61.611 & 26.238 \\ 
  bn180620354 & 08:29:22.735 & -72.842 & 5.855 \\ 
  bn180710062 & 01:29:21.269 & -49.933 & 13.542 \\ 
  bn180720598 & 14:21:26.039 & -29.189 & 10.000 \\ 
  bn180728728 & 17:29:11.437 & -15.219 & 10.040 \\ 
  bn180822423 & 10:08:32.522 & -5.898 & 2.803 \\ 
  bn180822562 & 13:30:29.570 & -128.070 & 7.513 \\ 
  bn180822562 & 13:30:29.570 & -118.178 & 6.344 \\ 
  bn180906988 & 23:42:36.388 & -2.471 & 1.039 \\ 
  bn180929453 & 10:52:35.121 & -1.456 & 0.606 \\ 
  bn181007385 & 09:14:19.608 & -23.373 & 3.996 \\ 
  bn181008877 & 21:04:29.161 & -131.183 & 27.879 \\ 
  bn181119606 & 14:32:19.202 & -2.566 & 1.798 \\ 
  bn181122381 & 09:09:04.964 & -1.937 & 0.299 \\ 
  bn181203880 & 21:06:37.705 & -6.482 & 0.870 \\ 
  bn181222279 & 06:42:52.975 & -79.631 & 40.808 \\ 
  bn190114873 & 20:57:02.490 & -5.573 & 1.942 \\ 
  bn190114873 & 20:57:02.490 & -2.854 & 1.537 \\ 
  bn190205938 & 22:31:11.876 & -40.086 & 9.198 \\ 
  bn190228973 & 23:21:30.204 & -15.148 & 7.989 \\ 
  bn190310398 & 09:33:20.756 & -49.157 & 4.120 \\ 
  bn190315512 & 12:17:39.138 & -366.193 & 6.912 \\ 
  bn190323879 & 21:05:17.600 & -893.855 & 26.624 \\ 
  bn190324947 & 22:44:18.392 & -17.146 & 2.474 \\ 
  bn190326314 & 07:32:13.823 & -27.769 & 1.672 \\ 
  bn190326314 & 07:32:13.823 & -18.099 & 2.115 \\ 
  bn190610750 & 18:00:04.042 & -14.819 & 1.160 \\ 
  bn190611950 & 22:48:51.696 & -62.594 & 20.082 \\ 
  bn190719624 & 15:00:01.045 & -86.830 & 1.579 \\ 
  bn190806675 & 16:12:34.836 & -1.664 & 1.188 \\ 
  bn190828542 & 12:59:58.210 & -46.588 & 38.536 \\ 
  bn190829830 & 19:56:40.582 & -47.965 & 5.565 \\ 
  bn190901890 & 21:21:37.555 & -63.144 & 20.014 \\ 
  bn190930400 & 09:38:17.809 & -162.830 & 40.308 \\ 
  bn191019970 & 23:18:48.942 & -96.333 & 29.779 \\ 
  bn191026350 & 08:23:43.801 & -5.943 & 4.110 \\ 
  bn191031025 & 00:39:28.692 & -178.171 & 10.422 \\ 
  bn191101895 & 21:28:37.561 & -44.664 & 1.903 \\ 
  bn191111364 & 08:44:52.025 & -25.765 & 16.425 \\ 
  bn191225309 & 07:26:50.763 & -94.689 & 2.024 \\ 

\end{longtable}
\twocolumngrid


\end{document}